# Exploring the transformation of user interactions to Adaptive Human-Machine Interfaces

Angela Carrera-Rivera[☼]
aicarrera@mondragon.edu
Mondragon Unibertsitatea
Mondragon, Basque Country, Spain

Daniel Reguera-Bakhache
dreguera@mondragon.edu
Mondragon Unibertsitatea
Mondragon, Basque Country, Spain

Felix Larrinaga
flarrinaga@mondragon.edu
Mondragon Unibertsitatea
Mondragon, Basque Country, Spain

Ganix Lasa
glasa@mondragon.edu
Mondragon Unibertsitatea
Mondragon, Basque Country, Spain

## ABSTRACT

Human-machine interfaces (HMI) facilitate communication between humans and machines, and their importance has increased in modern technology. However, traditional HMIs are often static and do not adapt to individual user preferences or behavior. Adaptive User Interfaces (AUIs) have become increasingly important in providing personalized user experiences. Machine learning techniques have gained traction in User Experience (UX) research to provide smart adaptations that can reduce user cognitive load. This paper presents an ongoing exploration of a method for generating adaptive user interfaces by analyzing user interactions and contextual data. It also provides an illustrative example using Markov chains to predict the next step for users interacting with an app for an industrial mixing machine. Furthermore, the paper conducts an offline evaluation of the approach, focusing on the precision of the recommendations. The study emphasizes the importance of incorporating user interactions and contextual data into the design of adaptive HMIs, while acknowledging the existing challenges and potential benefits.

## KEYWORDS

HMI, Adaptive user interfaces, recommendation systems

## 1 INTRODUCTION

Human-machine interfaces (HMIs) serve as cognitive and communicative platforms for humans and machines, facilitating the transmission of information [8]. Research on User Experience (UX) in this domain aims to enhance not only usability and performance but also overall user satisfaction, by considering the user's motivational goals during an interaction. According to Aranburu et al. [2], users are motivated by a sense of autonomy and competence. To enhance UX, manufacturing companies should consider the factors intrinsic to these motivators, such as the system's ability to facilitate intuitive and rapid task completion, as well as its capacity to anticipate user needs. The latter is closely linked to the user's cognitive capacity, affecting the memory required to complete a task [5].

Adaptation is defined as the capacity of systems to fit individual user physical and mental abilities, as well as the context of use and platform capabilities [20]. Since the early days of Human-Computer Interfaces (HCI), Adaptive User Interfaces (AUIs) have played a central role and have been considered a key characteristic of smart UX [12]. The use of AUI provides personalization, individually tailored instructions, and assistance, which cannot only improve the performance of the user but also enhance UX and acceptance of use.

Recent research has explored the analysis of operator behavior and prediction of future actions based on previous sequences of human-machine interactions. By analyzing the sequences of actions an operator takes, valuable insights can be gained into how operators interact with systems, and opportunities for improving performance and efficiency can be identified. This article presents a data-driven methodology to develop AUIs in the context of HMI, that involves data acquisition, pre-processing and the analysis of the interactions through a recommender system to predict the next action of an operator based on their previous interactions with a system. This approach can be used to develop AUIs that provide real-time personalized instructions and assistance to the user. To demonstrate the use of the methodology, we simulate the use of an industrial mixing machine and show how our approach can provide personalized and context-aware assistance to the operator.

The structure of this article is as follows: Section 2 begins with the fundamental background and related work, Section 3 explains the methodology used and each of its steps, and Section 4 provides an experiment that follows the methodology and simulates the use of an industrial mixing machine. Finally, Section 5 discusses the results, challenges and limitations and provides the conclusion.

## 2 BACKGROUND

An HMI represents the means of communication between humans and machines. These interfaces have become increasingly important in modern technology, allowing users to interact with complex systems and devices intuitively and efficiently. Over the years, HMIs have evolved significantly, from simple command-line interfaces to sophisticated interfaces [16] that can take many forms, such as graphical user interfaces, voice interfaces, touchscreens, and augmented reality displays. However, traditional HMIs are often static not adapting to individual user preferences, needs, or contexts. The lack of adaptivity can result in a high learning curve, reduced efficiency, and poor UX. This is precisely where AUI can prove helpful.

An AUI is considered as an *artifact* that can alter the interface characteristics and functionality based on prior experience with users in a dynamic form by understanding user sequential behavior. Sequential behavior can be an indicator of the level of expertise or familiarity with a given device or machine. In the literature, multiple authors have studied user actions to understand tasks and user goals in various fields. One example from engineering design is the work of McComb et al. [11] who presented an approach to assist designers with previous design iterations using sequence learning. In assembly assistance, Gellert and Zamfirescu [7] described the development of an automated assembly training station for manual operations, which suggests the next assembly step through Markov chain predictors. Wallhoff et al. [22] proposed an adaptive framework that utilized gesture and object recognition devices to guide operators in complex assembly tasks.

In the context of AUI, several studies have focused on industrial setting. Neira et al. [14] focused on developing AUI for Android devices based on operators, roles, and system states in manufacturing scenarios. Additionally, Mourtzis and Vlachou [13] presented an adaptive Cyber Physical System (CPS) to improve factory performance by assisting operators in decision-making through adaptive scheduling. Finally, AUIs were found to facilitate users in complex tasks in Zhang et al. [26]. Several of these approaches are model-based and do not consider in particular user interactions. Methods to analyze human-machine interactions exploit machine learning techniques, such as recommendation systems to provide adaptation and personalization for users, which can help reduce the cognitive load and thus increase the positive feelings of the user[4, 5]. A recommendation system is a set of tools (machine learning, algorithms, and data analysis) whose purpose is to support users in the decision-making process[5]. In particular, a sequence-aware recommendation system characterizes by having an ordered and/or timestamped list of past user actions as input [17] and the output is the ranked list of recommended elements or actions. However, sometimes the order of these recommendations can be critical, and users would prefer to follow the provided order, which is often the case when providing user assistance.

## 3 METHODOLOGY

In this section, we present the methodology followed in this article, which is depicted in Figure 1. The goal is to assist operators in their interactions with a machine by predicting their next steps when using an HMI. Each of the stages will be described next.

### 3.1 Data acquisition

Acquisition of human-machine interactions is a process that will vary according to how the user interacts with the machine. Interactions are usually linked to visual elements of the interfaces (i.e. buttons, tabs, knobs). Automated logging of user interactions was chosen as the approach for data acquisition in our methodology. This approach has several advantages over other data collection methods, including being non-intrusive, capturing data in real-time, and reducing errors associated with manual data entry. Automated logging can be achieved using specialized software such as Matomo or Google Analytics for web or mobile apps, or by building custom logging tools directly onto the system. In some cases, automated logging can be implemented at the interface level without requiring access to the system's internals. However, this approach can be challenging, especially for complex interfaces, and may require access to the underlying code of the system being studied, which is not always possible. For example, Zhou et al. [27] presented an approach for monitoring back-office staff using a screen-mouse-key logger that captured images, mouse, and key actions, along with timestamps, and was transformed into a UI log through image-analysis techniques.

However, capturing user interactions can also raise privacy concerns if personal data is inadvertently captured or if users are not aware that their interactions are being logged. Additionally, the use of machine data is a valuable source of contextual information at the moment to provide the adaptation [4], since alarms, status, and measures of sensors and actuators can potentially change how the user will interact with the machine.

### 3.2 Pre-processing

The purpose of the step is to generate valid sequences of user interactions based on the process or task to fulfill. A sequence of events $E = <e_1, e_2, ..., e_m>$ ($e_i \in D$) is represented as a list of ordered events $e_i$, where D is a set of known events, and the order is defined by $i$. In other words, $e_i$ occurs before $e_{i+1}$. In this work, for a sequence to be considered valid, it must contain at least two events [18]. Using this definition and the raw interactions as input, valid interaction sequences can be defined as $s_i = [e_{begin}, e_1, ..., e_k, e_{end}]$, where $s_i$ is a set of events, and:

- The events $e_{begin}$ and $e_{end}$ are known, indicating the beginning and end of the interaction sequence.
- The length of the interaction sequence is determined by the variable $l$, which should be $>= 2$.

As previously stated it is important to analyze the process or task and define a goal to correlate elements of the HMI with the goal completion. For example, one could determine that a user has completed a goal when they click the "accept" button. The *"Valid sequences extractor"* algorithm presented by Reguera-Bakhache et al.[18] is used to extract the sequences.

### 3.3 Recommender system

UX is the result of the user's internal state, the characteristics of the designed entity, and the context features involving the interaction [3]. Therefore, contextual factors should be considered in recommendation systems to provide more personalized and targeted recommendations that better align with the user's preferences and needs. Context-Aware Recommender Systems (CARS) have been proven to provide better customization and an improvement in recommendation accuracy [23]. Contextual pre-filtering is a common technique used in CARS, where contextual attributes such as sequence order, time, and user demographics are identified to filter and prioritize data used in the recommendation system's algorithms [10]. Then, an analysis of user sequences of past interactions is performed to predict the *next step* for an operator. To this end, we employed the Markov model, which is widely used for sequential recommendations [11, 19, 24] and can be a starting point for the creation of models such as hidden Markov model[1]. A Markov chain is a stochastic model that transitions between a finite number of possible states [21]. In a first-order Markov chain, it is assumed that the probability of the current state will only depend on its immediate previous state $P(X_t | X_{t-1})$, also known as the *Markov property*. It is also possible to create a higher-order Markov chain that can consider $n$ previous states $P(X_t | X_{t-1}, ..., X_{t-n})$, thus modeling a degree of "memory" within the system [11]. In sequence-aware recommendation systems, the Markov property assumes that the ensuing user interactions only depend on a limited number of the most recent preceding actions [17], which for UX analysis can highlight users' behavior patterns. The next step is determining the appropriate number of recommendations to provide to the user, commonly referred to as the "top-k", in which the order of options provided is also important since the first option presented should be the most adequate as the prediction in the operator´s next step. Too many options can be non-beneficial to the user's mental load, leading to reluctance towards UI adaptation.

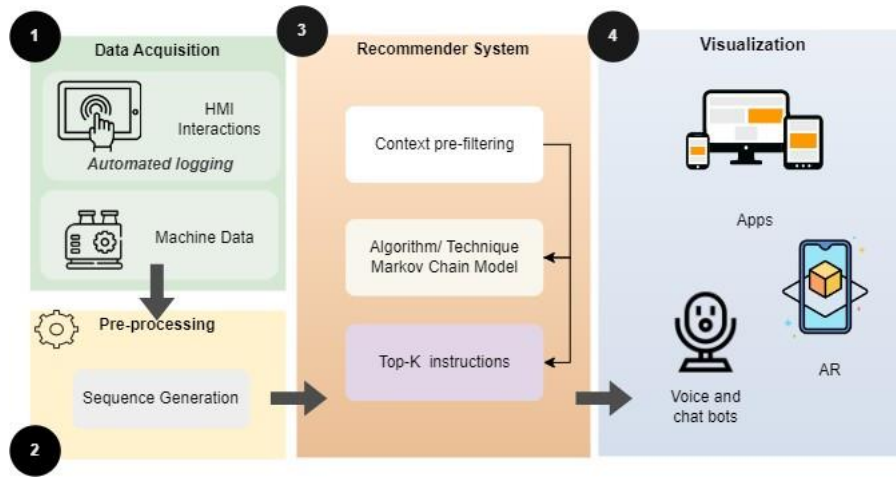

**Figure 1: Methodology for AUI**

### 3.4 Visualization

Providing the visual elements in which the assistance or adaptation of the HMI is going to occur is a critical aspect of the user interface design and is the closest to the UX. The appropriate visualization technique depends on the type of recommendation being made and the context of the user's interaction. For example, in mobile and web apps, graphical elements such as messages, icons, and buttons can be used to represent recommended options. In augmented reality, virtual objects can be overlaid onto the user's physical environment to indicate recommended actions or items. In natural HMIs, chatbots and voice bots with natural language generation techniques can be used to provide personalized recommendations in a conversational format. Overall, it is important to ensure that the visualization does not overwhelm the user or interfere with the overall user experience.

In this work, we focus on graphical HMIs, which remains relevant in the present context. Oestreich et al. [15], in their study, highlight the significance of clearly defining the target of adaptation, which in this case is oriented towards the instructions required to complete a process. The adaptation of assistance systems involves two critical aspects: structure and content. The structural part encompasses the sequence of instructions necessary to fulfill a process. To achieve this, a recommender system generates predictions for the next instruction, which are then communicated and executed through the HMI. On the other hand, the content is presented through text and messages overlaid on the app. These textual elements serve as informative cues, providing relevant information and directions. The goal of visualization is to optimize the presentation of this content to enhance the user experience and improve the efficiency of task completion.

## 4 EXPERIMENTATION

For this experiment, a web app was created in Next.js, an open-source framework based on React. The app simulates the use of a mixing industrial machine in the food sector, as presented in Figure 2. For ***data acquisition***, interactions were captured from a total of 24 users who consent to have their interactions with the app recorded through a built-in component for automatic logging. It is important to mention that every element in the HMI has a unique ID that allows for tracking of elements. In total, 10,608 interactions were captured. During the ***pre-processing*** stage, a total of 1,358 valid interaction sequences were generated.

For the ***recommendation system***, we used the role of the user and shift as sources of context (Figure 2), as well as the order of previous interactions, to adapt the interface. As described before, we employed the Markov model for sequential recommendations. The states on our model represent each interaction with the elements of the designed HMI. Thus, the set of states contains all possible sequences of user interactions. These sequences are represented as vectors $e_1, e_2, ...., e_n$ of size $n$ where $e$ represents each element that the user interacted with.

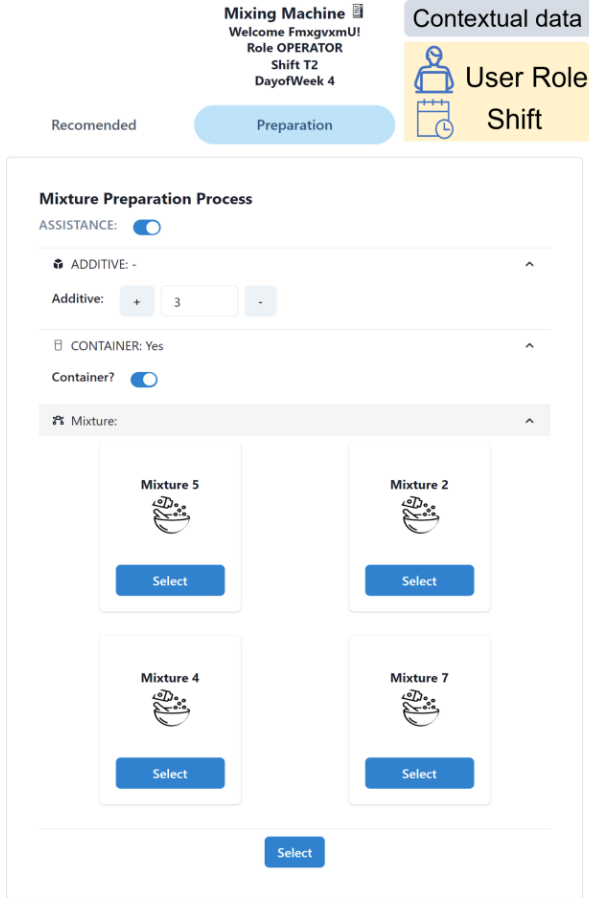

**Figure 2: Web app developed for the experiment**

The Markov chain model is defined by the values of the elements in its transition matrix **T** Element $T_{i\,j}$ in the matrix defines the probability of the next element will be $j$ given that the previous was $i$. This can be visually represented in a directed graph (Figure 3) where vertex are elements in the UI and edges are the state transitions. Training higher-order Markov chains follow a similar pattern as first-order Markov chains. However, each state is no longer represented by one single element and instead is $n$ tuples of elements where $n$ represents the order of the chain. Hence, a dependency with previous states exists in this scenario. First, second and third-order Markov chains were trained using the dataset, which was split. The last interaction of each user was considered for testing and removed from the training dataset. Each user has an average of 50 sequences and the sequence length has a median of 8 interactions. An incremental sequential evaluation was then performed, which consist of predicting the next item given the first $N$ elements of the sequence [17]. As presented in Figure 4, each sequence list was incrementally segmented into two vectors: *user selection* and *ground truth* which were used for performance evaluation. This process is repeated for each sequence in the testing dataset, where the number of iterations is determined by the length of the sequence, as outlined in Algorithm 1.

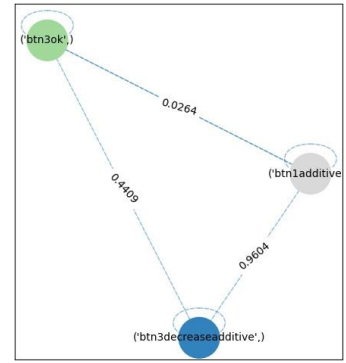

**Figure 3: First-order Markov chain state transition from a subset of HMI interactive elements**

In the context of recommendation systems, only a set of top $K$ options will likely be recommended. *Precision* and *Recall* at $K$ are commonly used metrics to evaluate recommendation systems, where precision represents the ratio that a recommended item at $K$ is on the set of overall recommendations, and recall represents the ratio of relevant elements present over the total of relevant elements [9]. On the other hand, in this scenario is important that the more relevant option for the user is presented first, where the use of Mean Reciprocal Rank (MRR) can evaluate systems that return a ranked list of elements. Therefore, as performance measures, we used the standard *Precision* and *Recall* at $K$ = 3, and MRR computed over each element of the test sequence. The performance scores achieved by the compared Markov chain models are summarized in Table 1.

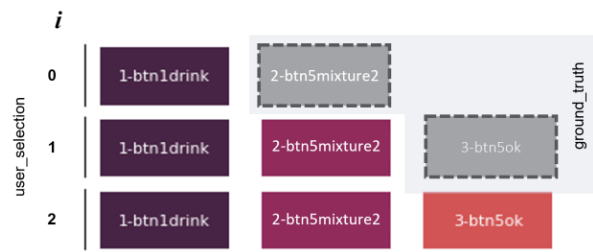

**Figure 4: Incremental sequence evaluation**

Table 1 shows that the precision increases for higher order Markov chains (∼ 5% increase), whereas the MRR increment is

**Algorithm 1** Incremental Sequence evaluation

**Require:** Set of test sequences $T = \{seq_1, ..., seq_n\}$ $seq_n = \{e_1, ..., e_n\}$
$topK \leftarrow 3$
  for $seq \in T$ do
    for $i = 1; i < seq.length; s++$ do
      $user\_selection \leftarrow seq[0 : i]$
      $ground\_truth \leftarrow seq[i : i+1]$
      /* Get an ordered list of recommended next interactions, get topK elements */
      $list\_recommendation \leftarrow MarkovModelRecommender(user\_selection)[0 : topK]$
    end for
  end for

| Metric | 1st.Order | 2nd.Order | 3rd.Order |
|---|---|---|---|
| Precision | 0.326605 (0.045337) | 0.342766 (0.075853) | 0.362626 (0.077289) |
| Recall | 0.900249 (0.138952) | 0.890848 (0.139525) | 0.887999 (0.142728) |
| MRR | 0.801013 (0.139366) | 0.814988 (0.146211) | 0.814724 (0.146954) |
| F1 | 0.479316 | 0.495053 | 0.514961 |

**Table 1: Mean Metrics results for each Markov chain model (with standard deviation in brackets)**

negligible (~1% increase), which can be explained as the users finding more pertinent UI elements in the sequence recommendation but not necessarily in a higher rank. The results suggest that the MRR would not improve more after the third-order. But overall, the model obtains a >80% in MRR, which suggests the best option of the sequence was ranked first.

However, performance metrics do not evaluate how user behavior is influenced by the system's suggestions or what percentage of recommendations are accepted by users. To achieve this systems should be tested in real-time with users, and compare the performance of both metrics and user feedback.[19].

"The final stage of the experimentation process involves the visualization component. Within the mobile app, contextual data from the user is detected and their last instructions are stored. The app establishes a connection with the Recommendation System component, as illustrated in Figure [insert figure number], which filters previous interactions based on the received context. These interactions are then utilized to generate predictions using second-order Markov chains.

The last part of the experimentation is the **Visualization**. Within the mobile app, contextual data from the user is detected and their last instructions are stored. The app establishes a connection with the Recommendation System component, as presented in Figure 6, which filters previous interactions based on the received context. These interactions are then utilized to generate predictions using second-order Markov chains. The top prediction is returned in JSON format. This means that the app keeps updating the last 2 user interactions as input of the algorithm that predicts which HMI interactive element has been most likely executed in the past as next instruction.

The moment of adaptation in this case occurs between the current and next instruction step. Since the last made interactions from the user trigger the execution of the next interactive UI element, which in this case is achieved in the app by using JavaScript.

According to Oestreich et al. [15], the user's control over adaptation has three dimensions to consider: automation, visibility, and participation. In this case, the app automates the execution of the interactive elements predicted, but using a toast message, we make it visible to the user what is occurring (Figure 5). The user participates in the adaptation by accepting the process or deactivating the assistance. It is important to communicate to the user since the goal of adaptation is to improve the UX, and confusion or frustration can decrease the overall acceptance of the system.

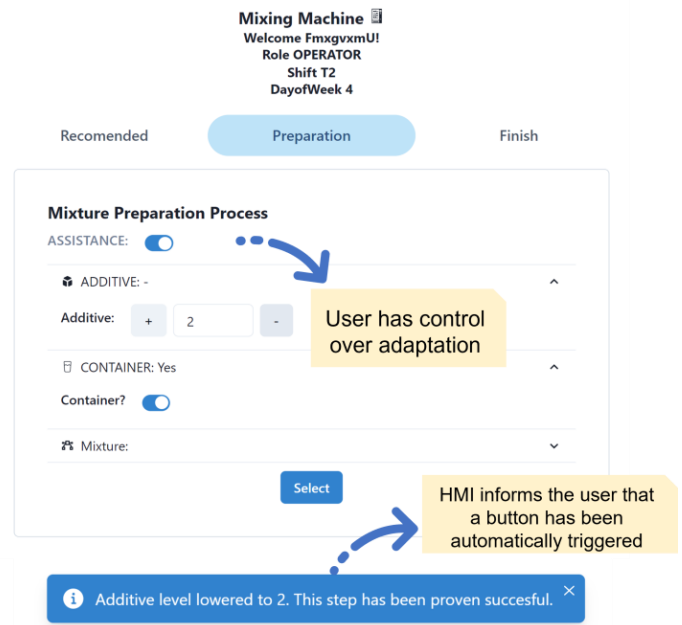

**Figure 5: Adaptation presented in the HMI**

## 5 CONCLUSION

This paper has highlighted the importance of incorporating adaptation in HMI to provide personalized experiences for users, in often static user interfaces.

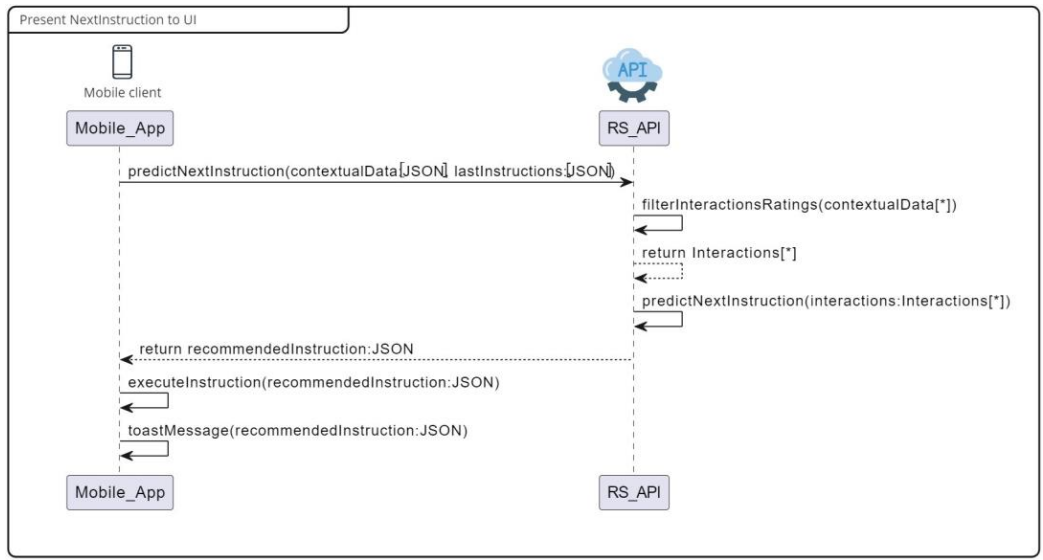

**Figure 6: Sequence Diagram of Mobile App Client and Recommendation System component**

By utilizing machine learning techniques, this work demonstrates the usage of user interactions and contextual data to provide assistance to operators on adaptive HMIs. The approach builds upon prior research that suggests the use of Markov Models for generating sequential predictions, and the results of our experiment showcase its potential for deployment in various industrial settings.

It is important to note that further validation through testing with users is necessary to establish the full effectiveness of this approach. In particular, future evaluations of the HMI should consider the impact of interface adaptations on end-user satisfaction [6], incorporating a combination of machine learning metrics, user feedback, and usability metrics such as click-through rate and time-on-task [25].

Moreover, there are challenges presented, such as the cold-start problem in cases where there is insufficient data to generate predictions and data sparsity that can arise in the context pre-filtering approach. Hybrid methods that combine multiple recommendation techniques are planned in future work to address these challenges.

The method presented emphasizes the significance of considering both user and machine data in HMI adaptation, which was not implemented in this particular experiment. But, we consider that there is a need for further research into approaches that effectively integrate these two data sources.

Overall, the study provides valuable insights into the potential of user interactions and Markov Models for adaptive HMIs, and presents opportunities for future research in this field.

## 6 ACKNOWLEDGMENTS

This project has received funding from the Department of Education, Universities, and Research of the Basque Government under the project Ikerketa Taldeak and the European Union's Horizon 2020 research and innovation programme under the Marie Sklodowska-Curie Grant No. 814078. Daniel Reguera-Bakhache is part of the Intelligent Systems for Industrial Systems research group of Mondragon Unibertsitatea (IT1676-22), supported by the Department of Education, Universities and Research of the Basque Country. Angela Carrera-Rivera and Felix Larrinaga are part of the Software engineering research group (IT1519-22).